\documentclass[aps,prb,showpacs,superscriptaddress,twocolumn]{revtex4}

\usepackage{graphicx,amssymb,amsmath,color}

\begin{document}
\title{Fingerprints of Majorana fermions in current-correlations measurements from a superconducting tunnel microscope}
\author{P. Devillard}
\affiliation{Aix Marseille Univ., Universit\'e de Toulon, CNRS, CPT, Marseille, France}
\author{D. Chevallier}
\affiliation{Department of Physics, University of Basel, Klingelbergstrasse 82, CH-4056 Basel, Switzerland}
\author{M. Albert}
\affiliation{Universit\'e C\^ote d'Azur, CNRS, Institut de Physique de Nice, France}

\begin{abstract}
We compute various current correlation functions of electrons flowing from a topological nanowire to the tip of a superconducting scanning tunnel microscope and identify fingerprints of a Majorana bound state. In particular, the spin resolved cross-correlations are shown to display a clear distinction between the presence of a such an exotic state (negative correlations) and an Andreev bound state (positive correlations). Similarity and differences with measurements with a normal tunnel microscope are also discussed, like the robustness to finite temperature for instance.
\end{abstract} 
\maketitle

\section{Introduction}

\begin{figure} 
\includegraphics[width=1\linewidth]{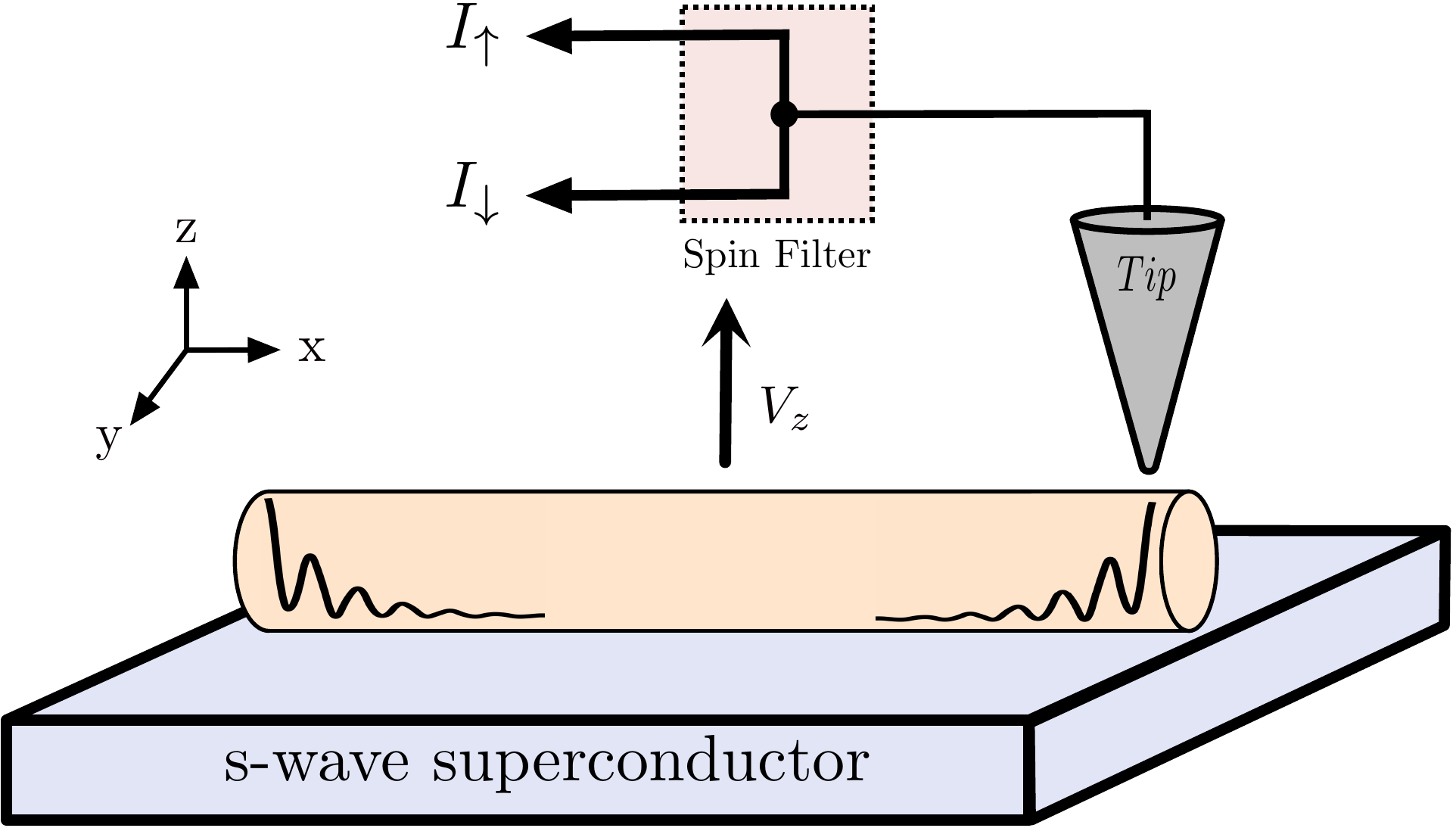}
\caption{(Color online) The system consists of a grounded TN driven in the topological phase (i.e. $V^2_z>\Delta_s^2+\mu^2$) carrying MBSs at its ends and in proximity with a biased SC STM tip. This tip is approached above the MBS allowing us to detect the noise as well as the spin current correlations of the current flowing between them. Note that the spin current correlations can be extracted using a spin filter along the output signal. In our study, we are deep in the topological phase meaning that the magnetic field is sufficiently large to polarize all the spins of the TN in its direction.}
\label{fig:Fig1}
\end{figure}

A Majorana bound state (MBS), in condensed matter physics, is a zero-energy quasi-particle with the specificity of being its own antiparticle \cite{Majorana}. Among many interesting properties, this exotic particle may, in particular, belong to the family of anyons \cite{Wilczek,ElliottFranz,Beenakkerprl} and therefore have a non-Abelian statistics which make it a very interesting object for quantum computation \cite{Aliceareview,Nayakreview}. Such states can be realized in various solid state systems and dimensionalities \cite{Beenakkerreview,Lutchyn}, the simplest one being the so-called Topological Nanowire (TN) which consists in a Rashba nanowire on top of an s-wave superconductor and in the presence of an external magnetic field \cite{Beenakkerreview,MourikKouwenhoven,Yacoby,Lutchyn,Oreg,Aliceareview,Szumniak,Rainis2013} as sketched on Fig. \ref{fig:Fig1}. Such a system can be tuned in the topological phase by choosing properly some experimental parameters such as the chemical potential of the superconductor or the external magnetic field \cite{Kitaev,FuKane}. A similar system, matching perfectly the TN but with more experimental degrees of freedom, has been recently developed with a chain of magnetic atoms deposited on top of a superconductor \cite{Pawlak,Shiba, Yazdani}.

Experimentally the presence of a MBS could be probed via the presence of a zero bias conductance peak (ZBCP) in the differential electrical conductance with a very specific quantized value of $2e^2/h$ \cite{PotterLee}. Unfortunately, the ZBCP is still far from being an unambiguous signature of a MBS since such a peak could originate from other phenomena such as Andreev bound states \cite{Shiba,Kellsetal}, weak anti-localization \cite{Pikulinetal}, disorder \cite{PotterLee,BagretsAltland} or Kondo resonances \cite{Sasakietal,AkhmerovBeenakkerZBCP,LeeJiangHouzet,Nilssonetal}. Moreover, the temperature being also important in the state of the art experiments, is an additional source of pollution in the sense that it suppresses the amplitude of the ZBCP and tends to blur the signal. This aspect has been studied recently leading to the conclusion that a superconducting STM tip would allow us to measure the signal while getting rid of the temperature broadening and therefore obtain cleaner signatures in the conductance peak and  Majorana wave function tomography \cite{Glazman,ChevallierKlinovaja,Setiawan2017_1,Setiawan2017_2}. However, spurious sub-gap states may still exist and contribute to smudge the signal coming from the MBS. This is why the community is still making efforts to find experimentally a smoking gun able to distinguish between the presence of MBSs and other sub-gap states such as ABS \cite{ChevallierAlbertDevillard,Valentini2016,Tanaka2010} or Kondo resonances \cite{AkhmerovBeenakkerZBCP,LeeJiangHouzet}.

A possible lead is to use very peculiar properties of Andreev reflection in the presence of a MBS. In that case, there exists a specific spin direction, called the Majorana polarization, along which electrons with positive spin projection are perfectly Andreev reflected (as holes with the same spin orientation) while the others are perfectly specularly (directly) reflected (as electrons with the same spin orientation). This specific reflection is called spin-selective Andreev reflection (SESAR) and can occur only in the presence of the MBS \cite{MITgroupPaLeeSesar,Fidkowskietal}. This will of course have important consequences on standard observables in mesoscopic physics like the noise or more generally current correlations functions \cite{Valentini2016,Landauer,Buttiker2000,Haim,Zazunov2016,Jonckheere2017} which have been shown to be very instructive, for instance, for probing the fractional charges in the quantum Hall effect at filling factor $\nu=1/3$ \cite{SaminadayarEtienneGlattli,Heiblum}. In our case of interest, it has been suggested that, even if the signal is noiseless, a spin filter can be used to study the spin current correlations between a normal metal (N) tip and the TN \cite{Lambert,Takane,AnantramDatta}. Using a properly oriented spin filter, the spin current correlations are always negative for a MBS and always positive or zero for an ABS.

In this paper, we study the current correlations coming from the detection of the MBS by a superconducting (SC) STM tip. By grounding the wire and applying a finite bias to the tip such that the quasi-particles injected in the wire are electron-like (the bias has to be of the order of the superconducting gap of the tip), we can make a clear comparison with the previous results obtained in the case of a normal metal tip \cite{Haim}. We calculate the zero frequency noise when the superconducting STM tip is placed just above the MBS and show that this noise is finite in contrast to what happens for the detection with a normal metal tip. Then, we calculate the spin current correlations and show that these correlations are negative in strong contrast with the detection of an ABS where the correlations are positive. We also point out that these correlations depend strongly on the ratio between the tunneling for spin up and spin down (this ratio can be changed by putting a magnetic barrier or applying an external electric field at the interface). We provide numerical results with the full range of possible tunneling from fully polarized to unpolarized situations. This opposite sign of correlations between ABS and MBS is thus a clear signature of their intrinsic difference which can be measured experimentally with the usual tools in current laboratories.

The article is organized as follows: in Sec. II, we describe the model under investigation. The third and fourth sections explain the calculation and discuss the results for, respectively, the noise and the spin current correlations of the current flowing between the tip and the TN. We finally conclude in the last section and give some insights of possible experiments. Technical details are discussed in the appendixes.

\section{Model} 

We consider a TN which consists of a Rashba nanowire in proximity with an s-wave superconductor of chemical potential $\mu$ and superconducting gap $\Delta_s$ in the presence of an external magnetic field $V_z$ applied along the $z$ direction as sketched in Fig. \ref{fig:Fig1}. Such a system experiences MBSs at its ends when tuning it in the topological phase ($V^2_z>\Delta_s^2+\mu^2$). For convenience, we suppose that the wire is sufficiently long so that the overlap between the two MBSs is negligible. The Bogoliubov-de Gennes (BdG) Hamiltonian of the TN along the $x$ axis reads 
\begin{eqnarray} \label{hamil_wire}
H_{TSC} = \frac{p_x^2}{2 m} \tau_z + i \alpha_R \tau_z\sigma_y \partial_x + V_z\sigma_z + 
\Delta_s \tau_x - \mu \tau_z,
\label{HTSC}
\end{eqnarray}
in the Nambu basis $(\psi^\dagger_{k,\uparrow}, \psi^\dagger_{k,\downarrow}, \psi_{-k,\uparrow},\psi_{-k,\downarrow})$ where $\Delta_s$ is the superconducting gap induced by proximity effect in the nanowire, $\mu$ is the chemical potential which is taken as the origin of energies and $\alpha_R$ is the Rashba coupling strength along the $y$ direction. Here, the $\tau_i$ ($\sigma_i$) denote the Pauli matrices acting respectively in the particle-hole (spin) space. A superconducting STM tip is approached above the TN and can be moved along the wire to perform its tomography and, meanwhile, detect the MBSs \cite{Yazdani,Glazman,ChevallierKlinovaja}. The Hamiltonian of the STM tip is the usual BCS Hamiltonian with s-wave pairing
 \begin{equation}
 H_{tip} = \sum_{{\bf k},\sigma} \epsilon_{{\bf k},\sigma} c_{{\bf k},\sigma}^{\dagger} c_{{\bf k},\sigma} + 
\Delta  c_{{\bf k},\sigma}^{\dagger} c_{{\bf- k}, -\sigma}^{\dagger} + h.c.,
\end{equation}
 where $\Delta$ is the superconducting gap of the tip. The tunneling Hamiltonian allowing the transfer of particles between the TN and the tip reads 
 \begin{equation}
   H_t = \sum_{\sigma} i t_{\sigma} \gamma \psi(0)  c_{\sigma} + h.c.\, 
\end{equation}
with $t_{\sigma}$ the hopping amplitudes taken to be real \cite{footnoteta}, $\psi(0)$ the Majorana wave function just below the tip taken at the end of the wire where the Majorana wave function is maximal and $\gamma$ the corresponding Majorana operator.  The hopping amplitudes $t_{\sigma}$ depend on the spin polarization of the MBS which can be calculated via Eq. (\ref{hamil_wire}) using the BdG equations \cite{Lutchyn,MITgroupPaLeeSesar,Prada}.
For the rest of the paper, we take $t_{\sigma}$ positive and set $\lambda \equiv t_{\downarrow}/t_{\uparrow}$. We neglect the contribution from the continuum of the TN above $\Delta$ because we bias the junction such that the quasi-particles from the tip flow into the MBS and not into higher subbands. 
 
\section{Noise} 

The absorption current noise at finite frequency reads 
\begin{equation}
S_a(\omega) = \int_{- \infty}^{\infty}
 \Bigl\langle (I(0)- \langle I \rangle ) (I(t) - \langle I \rangle)\Bigr\rangle e^{i \omega t} \, dt,
 \end{equation}
 where $I(t)$ is the total current operator and $\langle I \rangle$ is the average current. For the sake of clarity, we choose to use this non-symmetrized version and drop the subscript $a$ in the rest of the paper. It is straightforward to express the current in terms of the components of the reflection matrix ${\hat r}$ (its components 
read $r^{ee}$, $r^{hh}$, $r^{eh}$ and $r^{he}$, and are themselves two by two matrices in spin space) \cite{Nilssonetal}.
The broadening of the MBS due to the tunneling can be written as $\Gamma \equiv 2 \pi \nu_0 (\vert t_{\uparrow}\vert^2 + \vert t_{\downarrow}\vert^2)$, where $\nu_0$ is the density of states of the tip in the normal state. The ${\hat r}$ matrix is obtained via the Lippmann-Schwinger equation \cite{MahauxWeidenmuller,IidaWeidenmullerZuk}
\begin{equation}
 {\hat r} = I - 2 i \pi W^{\dagger}(E + i \pi W W^{\dagger})^{-1}W,
 \end{equation} 
 where $I$ is the identity matrix and $W$ the is ``contact'' matrix. When the system is in the topological phase with MBSs at the ends of the wire, $W$ reads 
\begin{eqnarray}
W_{\textrm{MBS}}= \sqrt{\nu_0} \begin{pmatrix}
\sqrt{\rho_+} t_{\uparrow},
\sqrt{\rho_+} t_{\downarrow},
\sqrt{\rho_-} t_{\uparrow},
\sqrt{\rho_-} t_{\downarrow}
\end{pmatrix},
\end{eqnarray}
 where $\rho_{\pm}=\rho(E\mp eV)$ is the dimensionless density of states for the electron-like (hole-like) quasi-particles normalized by $\nu_0$. When the applied voltage is slightly above the gap of the tip, so that $0 \leq eV-\Delta \ll \Gamma$, the following approximation can be made $\rho_{+}\simeq \bigl(\frac{\Delta} { 2}\bigr)^{1/2}\,(eV - \Delta - E)^{-1/2}$. For $\lambda=1$, the Fano factor is ${\rm F} = \frac{S(0) }{e \langle I_1 \rangle} \, = \frac{18 \pi - 56 }{ 12 - 3 \pi} \simeq 0.213$ (see Appendix A), where the total average current is denoted by $\langle I_1 \rangle$ and has been previously calculated in Ref. \cite{Glazman}. Thus, the noise is finite when using a SC tip in contrast to the case of a normal metal tip, where the noise vanishes. The physical explanation  is because the electrons with energies close to $\Delta$ are Andreev reflected with a probability not equal to one.  This gives rise to a suppression factor $4 - \pi$ in the conductance smaller than the usual quantized value $2 e^2/h$ \cite{Glazman,ChevallierKlinovaja}. To have a better understanding, let us put this in contrast with a normal metal tip detection, where the flow of electrons from the tip into the MBS occurs via Andreev resonant tunneling, with a perfect probability, leading to a noiseless signal \cite{Glazman}.

\section{Spin current correlations} 

By using a spin filter along the $z$ axis (see Fig. \ref{fig:Fig1}), such as a T-junction connected to two polarized quantum dots \cite{Haim,Haim2}, the total current is split into its two spin components $I_{\uparrow}$ and $I_{\downarrow}$ and the correlations between them can be calculated $P_{\sigma \sigma^{\prime}} \equiv \langle \delta I_{\sigma} \delta I_{\sigma^{\prime}}\rangle$ with 
$\delta I_{\sigma} \equiv I_{\sigma} - \langle I_{\sigma} \rangle$. 
The average spin current and its correlations have been studied extensively in the literature \cite{Nilssonetal} 
\begin{eqnarray}
\langle I_{\sigma} \rangle  \, = \, \frac{e }{ h} \,\int_0^{eV - \Delta}
\bigl\lbrack 1 - \sum_{\alpha=e,h}\textrm{sgn}(\alpha){\cal R}_{\sigma\sigma}^{\alpha\alpha}\bigr\rbrack \, dE,
\end{eqnarray}
 and
\begin{eqnarray}
P_{\sigma, \sigma^{\prime}}\, = \,  \frac{e^2 }{ h} \,\int_0^{eV - \Delta}
\sum_{\alpha=e,h}
\Bigl\lbrack \delta_{\sigma,\sigma^{\prime}} {\cal R}^{\alpha\alpha}_{\sigma,\sigma^{\prime}}
 - {\cal R}^{\alpha\alpha}_{\sigma\sigma^{\prime}}{\cal R}^{\alpha\alpha}_{\sigma^{\prime}\sigma} \nonumber \\
 + {\cal R}^{\alpha {\overline \alpha}}_{\sigma\sigma^{\prime}}{\cal R}^{\alpha{\overline \alpha}}_{\sigma^{\prime}\sigma}
 \Bigr\rbrack
  \,  dE,
\end{eqnarray}
where $\textrm{sgn}(\alpha)=\pm1$ for electrons/holes and ${\cal R}^{\alpha\beta}_{\sigma\sigma^{\prime}} = \sum_{\sigma^{\prime\prime}}
 r^{\alpha e}_{\sigma,\sigma^{\prime\prime}} r^{\beta e \, *}_{\sigma^{\prime},\sigma^{\prime\prime}}$. As a sum rule, we can easily check that the noise, calculated in the previous section, is $\sum_{\sigma,\sigma^{\prime}} P_{\sigma, \sigma^{\prime}}$. We again focus our study on the low voltage regime where $0 \leq eV- \Delta \ll \Gamma$.
  
For a normal metal tip, it has been shown that $P_{\uparrow\downarrow}$ is negative \cite{MITgroupPaLeeSesar,Nilssonetal}. Even if the signal is noiseless, the spin current correlations are finite. Indeed, within this widely used model for the TN (see Eq. (\ref{hamil_wire})), with  $B_z$ in the $z$ direction and the Rashba axis in the $y$ direction, the MBS is spin-polarized along a direction $\hat{n}$ which lies for small $\alpha_R$ in the $(x,z)$ plane \cite{SticletBenaSimon}. $\hat{n}$ can be, in principle, computed by solving the BdG equations. For instance, if the MBS is polarized along $|\!\uparrow_z\rangle$ corresponding to the system being deeply in the topological phase where the external magnetic field is way larger than the superconducting gap, then, $t_{\uparrow}=1$ and $t_{\downarrow}=0$. More generally, both 
$t_{\uparrow}$ and $t_{\downarrow}$ are taken to be real 
and their ratio $\lambda$ depends on $\hat{n}$. The particular case $\lambda=1$ corresponding to $t_{\uparrow}= t_{\downarrow}$ leads to a  spin polarization of the MBS along the $|\!\uparrow_x\rangle$ axis which can be achieved when $V_z$ is slightly above the critical Zeeman potential $V^c_z=\sqrt{\Delta_s^2 + \mu^2}$ \cite{MITgroupPaLeeSesar,SticletBenaSimon,Prada}. In such a configuration, the detection with a normal metal tip leads to the Andreev reflection of the spins $|\!\uparrow_x\rangle$ without reversing their spin while the spins $|\!\downarrow_x\rangle$ are specularly reflected due to the SESAR effect. Because the spin filter separates the $|\!\uparrow_z\rangle$ and  $|\! \downarrow_z \rangle$ currents, the transmitted spins have to be decomposed along the $z$ quantization axis which yields $P_{\uparrow\downarrow}/e\langle I_1 \rangle= -\frac{1 }{ 4}$ \cite{Nilssonetal,Haim}.

In the case of interest, the detection with a SC tip, the shot noise is finite. This raises the question about the persistence or not of the negativeness of spin current correlations. We will now show that the answer is positive. For $\lambda=1$ and a bias $eV$ slightly larger than the gap of the tip $\Delta$, the energies of the incoming particles can be classified in two categories: the low energy ones corresponding to energies just above $\Delta$ and not too close to $eV$ and the high energy ones corresponding to energies close to $eV$. By setting $\epsilon \equiv \sqrt{\frac{1 - \eta }{ 1 + \eta}}$ with $\eta \equiv \frac{E }{ eV - \Delta}$, the configuration we are interested in, namely the high energy case, gives $\epsilon\approx0$ meaning that the Andreev reflection is suppressed (see Eq. (\ref{rmbs})) \cite{Glazman}. The low energy configuration, which is not the purpose of this study, gives on the other hand strong Andreev reflection. The properties of the electrons in the high energy case can be encoded, up to the first order in $\epsilon$, in the matrix $\hat{r}$ written in the Nambu basis with quantization axis for the spin along the $x$ direction 
\begin{equation}\label{rmbs}
 \hat{r}_{\textrm{MBS}}= \begin{pmatrix}
-1+2\epsilon & 0 & -2 \sqrt{\epsilon} & 0 \cr
0 &  1 & 0 & 0 \cr
 -2\sqrt{\epsilon}& 0 & 1 - 2\epsilon & 0 \cr
0 & 0 & 0 & 1
\end{pmatrix}.
\end{equation}

\begin{figure} 
  \includegraphics[width=1\linewidth]{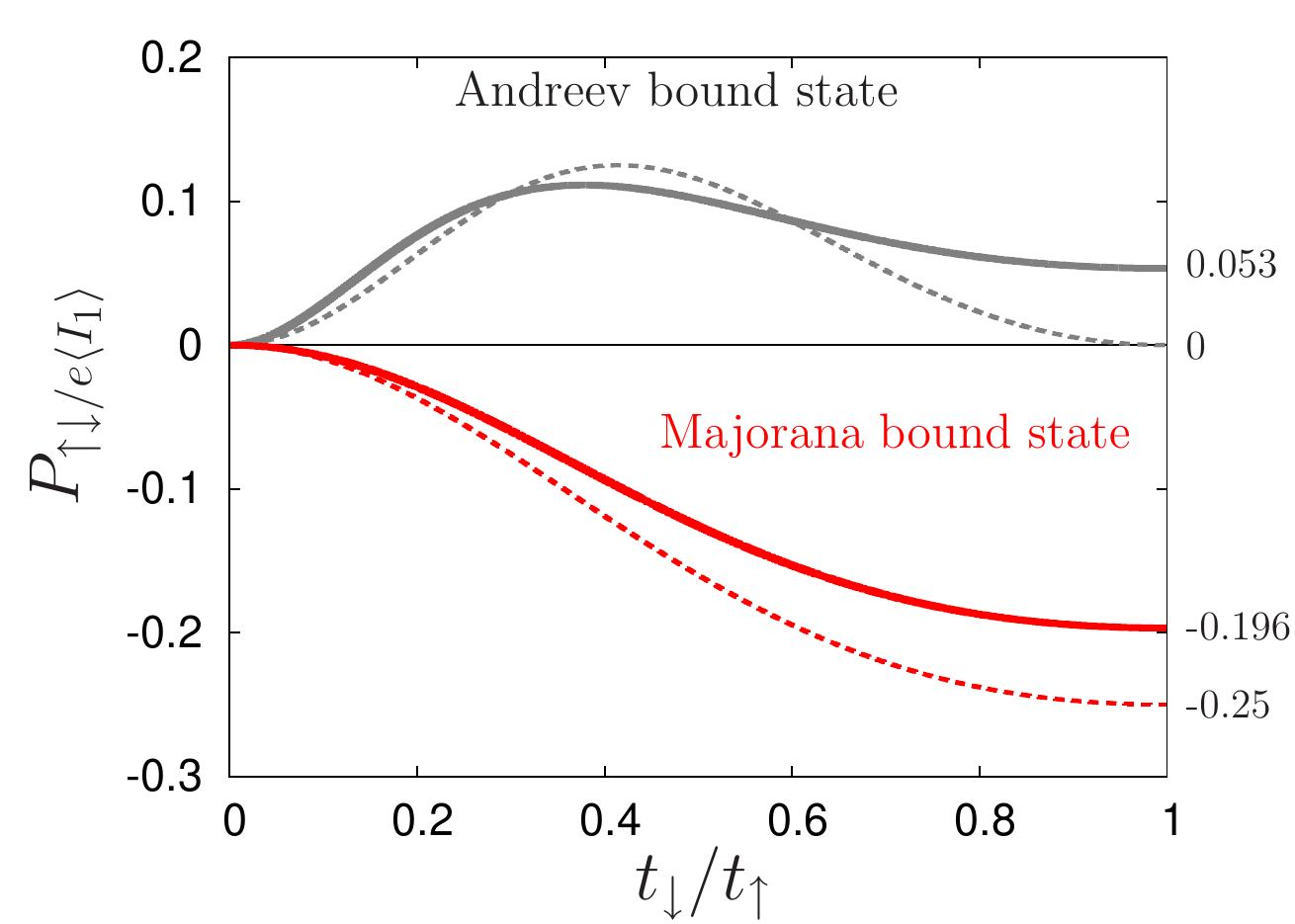}
  \caption{(Color online) Normalized spin current correlations ${\rm F}_{\uparrow\downarrow} \equiv P_{\uparrow\downarrow}/e\langle I_1 \rangle$ as a function of $\lambda$ for a MBS/ABS in the case of a SC STM tip (red/grey solid lines) and of a N STM tip (red/grey dashed lines).}
\label{fig:Fig2}
\end{figure}

In this basis, $|\!\!\downarrow_x \rangle$ electrons are perfectly specularly reflected and they are just spectators giving no contributions, neither to $\langle I_1 \rangle$ nor to $P_{\uparrow \downarrow}$. $|\!\uparrow_x\rangle$ electrons are mostly specularly reflected with amplitude $-1 + 2 \epsilon$ but a small amount of them are reflected via SESAR  with amplitude $- 2 \sqrt{\epsilon}$. This small contribution causes noise and generates positive contributions to $P_{\uparrow\downarrow}$. Note that this small component also alters $\langle I_1 \rangle$ through a suppression factor \cite{Glazman}. Therefore, in order to have a meaningful quantity, we introduce ${\rm F}_{\uparrow\downarrow} \equiv \frac{P_{\uparrow \downarrow} }{ e \langle I_1 \rangle}$ which can be seen as a Fano factor and where $P_{\uparrow\downarrow}$ is normalized by $\langle I_1\rangle$. The analytical calculation gives ${\rm F}_{\uparrow\downarrow} = - \frac{68 - 21 \pi }{ 48 - 12 \pi} \simeq -0.197$ (see Appendix B), slightly smaller than the $-1/4$ with a N tip for $\lambda=1$. On the contrary for $\lambda=0$, the current is fully polarized in the  $|\!\uparrow_z\rangle$ direction, thus leading to zero spin current correlations. In addition to these limiting cases, we have plotted $F_{\uparrow\downarrow}$ as a function of $\lambda$ in the general case $\lambda \not= 0,1$ both for a N tip and a SC tip on Fig. \ref{fig:Fig2}. One can clearly see that spin correlations remain negative in the presence of a MBS.

We now compare to a typical ABS case. The microscopic model of Eq. (\ref{hamil_wire}) exhibits ABS in the trivial phase below the transition, when $V_z<\sqrt{\Delta^2_{s}+\mu^2}$. In order to get a qualitative idea, we consider the tunneling between the tip and the ABS  such as
$H_t = \sum_k a^{\dagger} \Bigl(t_{\uparrow} \psi_{k,\uparrow} + t_{\downarrow}\psi_{k,\downarrow}^{\dagger}\Bigr)  + h.c.$, 
where $a$ stands for the ABS annihilation operator \cite{MITgroupPaLeeSesar}.  In this particular case, the ``contact'' matrix $W$ changes to
\begin{equation}
W_{ABS} = 
\begin{pmatrix}
\sqrt{\rho_+} t_{\uparrow} &0   &0 &\sqrt{\rho_-} t_{\downarrow} \cr
0 & \sqrt{\rho_+} t_{\downarrow} & \sqrt{\rho_-} t_{\uparrow}&0
\end{pmatrix}.
\end{equation}
For an ABS and independently of the type of tip, the usual Andreev reflection occurs and reverses any spin in the opposite one.  $P_{\uparrow \downarrow}$ can be written as
\begin{eqnarray}\label{cross_corre2}
P_{\uparrow\downarrow} = \frac{e^2 }{ h  } \int_0^{eV - \Delta} 
\Bigl\lbrack | r^{eh}_{\uparrow \downarrow} |^2  \bigl(1 -  | r^{eh}_{\uparrow \downarrow} |^2 \bigr)
 + (\uparrow \leftrightarrow \downarrow) \Bigr\rbrack \, dE\, ,
\end{eqnarray}
 which is similar to the noise of a QPC with transmission coefficient $T$ except that $T$ is replaced by the Andreev reflection coefficient. For a N tip and $\lambda=1$, there is perfect Andreev reflection for spins $|\! \uparrow_x\rangle$ and $|\! \downarrow_x\rangle$ which gives $P_{\uparrow\downarrow}=0$ (perfect transmission induces no noise). The same thing occurs for $\lambda=0$ where the spin along the $z$ direction is perfectly Andreev reflected and the other one is fully blocked. In all other intermediate cases ($\lambda\neq 0,1$), $P_{\uparrow\downarrow}$ has a finite value \cite{Haim,Haim2}. 

For a SC tip and $\lambda=1$, we can write $\hat{r}$ (with the quantization axis along $x$) in the ABS case
\begin{equation}
{\hat r}_{\textrm{ABS}}= 
\begin{pmatrix}
-1 + 2 \epsilon & 0 &  - 2 \sqrt{\epsilon} &0 \cr
0  & -1 + 2 \epsilon & 0 & 2 \sqrt{\epsilon} \cr
- 2 \sqrt{\epsilon}& 0 & 1 -2 \epsilon & 0  \cr
0 & 2 \sqrt{\epsilon} & 0  & 1 - 2 \epsilon 
\end{pmatrix}. 
\end{equation}
It is easy to see that both $|\!\uparrow_x \rangle$ and $|\!\downarrow_x \rangle$ are mostly specularly reflected but a small amount of them are reflected via SESAR with an amplitude $-2\sqrt{\epsilon}$ which generates a positive contribution to $P_{\uparrow\downarrow}$. The analytical result gives $F_{\uparrow\downarrow} = (9 \pi - 28)/(24 - 6 \pi)\simeq0.053$ (see Appendix B). 

To have a better view on N vs. SC tip and ABS vs. MBS, we have plotted in Fig. \ref{fig:Fig2}, $F_{\uparrow\downarrow}$ as a function of $\lambda$ for an ABS (grey lines) and a MBS (red lines) and in the case of a normal metal tip (dashed lines) and superconducting one (solid lines). In order to distinguish a MBS from an ABS, we propose to measure the sign of the spin current correlations of the sub-gap states via STM spectroscopy using a superconducting tip. Experimentally, the tunneling between the tip is generally supposed to be spin independent ($\lambda=1$) leading to a clear signature corresponding to the sign of these correlations. On top of that, we argue that, even if the response to the detection between a N tip and a SC one are very similar, the superconducting tip has the advantage to strongly reduce temperature effects which pollute the signal.

A legitimate question we can ask is about the spin decoherence issue for the electrons entering in the detection scheme, namely the SC tip plus the spin filter. Indeed, the spin decoherence time in superconductors is generally quite small (i.e. $\sim100$ps for Aluminium \cite {Charis}) which means that the measurement has to be shorter than this time. A quick calculation gives $\sim200\mu$m for the spin decoherence length with a Fermi velocity of $\sim2.10^6$ m/s leading to a reasonable system size allowing us to detect our effect.

\section{Conclusion} 

We have studied the noise and the spin current correlations when a superconducting STM tip is placed above a system hosting a MBS, to differentiate it from an ABS. First, we have shown that the noise gets a finite value for the detection using a superconducting tip coming from the non perfect Andreev reflection occurring at energy slightly larger than the superconducting gap. The detection using a normal tip leads to a well known result where the signal is noiseless due to the perfect transmission of the electron. A second result concerns the spin current correlations, we have found that the sign of these correlations are opposite for a current flowing into a MBS and an ABS giving us an opportunity to distinguish them via an STM measurement with a SC tip. The key advantage of using a superconducting tip is the possibility to get rid of the temperature effect because of the protection due to the gap \cite{ChevallierKlinovaja}. A future study would be to tackle more complicated cases such as ribbons with more than one MBS on each end \cite{SticletBenaSimon}. The second perspective is to study in detail the surface of a 3D topological insulator via superconducting STM spectroscopy and extract their properties depending on the symmetries they have (i.e. time reversal, etc.).

\section*{Acknowledgments} We would like to acknowledge helpful discussions with C. Dutreix and D.~I. Pikulin. The work of D. C. was supported by the Swiss NSF and NCCR QSIT.
 
\appendix

\section{Calculation of the noise in the case of a MBS detection}

The zero frequency noise for the total current reads
\begin{equation}
S \, = \, \langle (I - \langle I \rangle)^2 \rangle,
\end{equation}
 with $I = I_{\uparrow} + I_{\downarrow}$. In the spin independent case where $\lambda=1$, this equation can be simplified as
 \begin{equation}
 S = 2 (P_{\uparrow\uparrow} + P_{\uparrow\downarrow}).
\end{equation}
To extract the value of $S$, we need to calculate $P_{\uparrow\uparrow}$ and $P_{\uparrow\downarrow}$. Since we need to calculate the latter one for the spin current correlations, its calculation is derived in the first part of the Appendix B. It remains to calculate $P_{\uparrow\uparrow}$. From the Eq. (7) in the main text, the spin current correlations $P_{\uparrow\uparrow}$ can be expressed in terms of the reflection matrix $\hat{r}$ elements such as
\begin{eqnarray}
P_{\uparrow\uparrow} & = & \frac{e^2 }{ h} \int_0^{eV-\Delta} 
 ({\cal R}^{ee}_{\uparrow\uparrow}+{\cal R}^{hh}_{\uparrow\uparrow}
 + 2 {\cal R}^{eh}_{\uparrow\uparrow} {\cal R}^{he}_{\uparrow\uparrow} \nonumber \\ 
 & & - {\cal R}^{ee\, 2}_{\uparrow\uparrow} - {\cal R}^{hh\, 2}_{\uparrow\uparrow}) dE \\
& =& \frac{e^2 }{ h} (eV - \Delta) I_x, \nonumber
 \end{eqnarray}
 with $ {\cal R}^{x,y}_{\sigma,\sigma^{\prime}} \, \equiv \, 
\sum_{\sigma^{\prime\prime}} {\hat r}^{x,e}_{\sigma,\sigma^{\prime\prime}}
 {\hat r}^{y,e \, *}_{\sigma^{\prime},\sigma^{\prime\prime}},
$
where $x,y$ run for electron or hole and $\sigma$, $\sigma^{\prime}$ and $\sigma^{\prime\prime}$ are the spin orientations with quantization 
along the z axis. The integral $I_x$ is defined as $I_x= \int_0^1 u (3-u) (1+u)^{-2} dx$ where $u=\sqrt{1 - x^2}$. After integration, we get $I_x= \frac{5 \pi }{ 2} - \frac{22 }{ 3}$. For $\lambda=1$, the average current $\langle I_1 \rangle$ is 
\begin{equation}
\langle I_1 \rangle=\frac{2e}{h}(4-\pi)(eV-\Delta),
\end{equation} 
 which leads to 
 \begin{equation}
 F = \frac{S }{  e \langle I_{1}\rangle} \, = 
\frac{18 \pi - 56 }{ 12 - 3 \pi} \simeq 0.213.
\end{equation} 

\section{Calculation of the spin current correlations in the case of a MBS and an ABS detection}

From the Eq. (7) in the main text, the spin current correlations $P_{\uparrow\downarrow}$ can be expressed in terms of the reflection matrix $\hat{r}$ elements such as
\begin{eqnarray}\label{spincor_def}
  P_{\uparrow\downarrow} \, &=& \, \frac{e^2 }{ h} \int_0^{eV - \Delta}
  \Bigl\lbrack  - \, {\cal R}^{ee}_{\uparrow,\downarrow}  {\cal R}^{ee}_{\downarrow,\uparrow} 
  - \, {\cal R}^{hh}_{\uparrow,\downarrow}  {\cal R}^{hh}_{\downarrow,\uparrow} \nonumber \\
  & &  + {\cal R}^{eh}_{\uparrow,\downarrow}  {\cal R}^{he}_{\downarrow,\uparrow} 
  + {\cal R}^{he}_{\uparrow,\downarrow}  {\cal R}^{eh}_{\downarrow,\uparrow}\Bigr\rbrack \, dE,
\end{eqnarray}
with $ {\cal R}^{x,y}_{\sigma,\sigma^{\prime}} \, \equiv \, 
\sum_{\sigma^{\prime\prime}} {\hat r}^{x,e}_{\sigma,\sigma^{\prime\prime}}
 {\hat r}^{y,e \, *}_{\sigma^{\prime},\sigma^{\prime\prime}},
$
where $x,y$ run for electron or hole and $\sigma$, $\sigma^{\prime}$ and $\sigma^{\prime\prime}$ are the spin orientations with quantization 
along the z axis. By introducing the proper contact matrix into Eq. (4) of the main text, we can compute the spin current correlations and thus $F_{\uparrow\downarrow}$.  

\subsection{Majorana Bound State case}

In the most general case of complex hopping amplitudes, for a MBS, the reflection matrix ${\hat r}_{MBS}$ reads
\begin{eqnarray}\label{cont_mbs}
\begin{pmatrix}
\frac{R_+ \vert \lambda \vert^2 - R_- }{ {\tilde \Gamma}} & - 2 \frac{\rho_+ \lambda  }{ {\tilde \Gamma}} 
& - 2 \frac{\sqrt{\rho_+\rho_-} }{ {\tilde \Gamma}\varphi^2} & 
- 2 \frac{\sqrt{\rho_+\rho_-} }{ {\tilde \Gamma}\varphi^2} \lambda^* \cr
 - 2 \frac{\rho_+ \lambda^* }{ {\tilde \Gamma}} & \frac{R_+ - R_-  \vert \lambda \vert^2 }{ {\tilde \Gamma}} &
- 2 \frac{\sqrt{\rho_+\rho_-}}{ {\tilde \Gamma}\varphi^2}\lambda^* & 
- 2 \frac{\sqrt{\rho_+\rho_-} }{ {\tilde \Gamma} \varphi^2} \lambda^{*\,2} \cr
- 2 \frac{\sqrt{\rho_+\rho_-} }{ {\tilde \Gamma} \varphi^{-2}} & 
- 2 \frac{\sqrt{\rho_+\rho_-} }{ {\tilde \Gamma} \varphi^{-2}} \lambda & 
\frac{R_+ \vert \lambda \vert^2 + R_- }{ {\tilde \Gamma}} &  - 2 \frac{\rho_- \lambda^*  }{ {\tilde \Gamma}} \cr
 - 2 \frac{\sqrt{\rho_+\rho_-} }{ {\tilde \Gamma} \varphi^{-2}} \lambda  
&  - 2 \frac{\sqrt{\rho_+\rho_-} }{ {\tilde \Gamma} \varphi^{-2}} \lambda^2 &
 - 2 \frac{\rho_- \lambda  }{ {\tilde \Gamma}} &\frac{R_+ + R_- \vert \lambda \vert^2 }{ {\tilde \Gamma}}
\end{pmatrix},
\end{eqnarray}with $R_{\pm} \, \equiv  \rho_+ \pm \rho_-$, 
${\tilde \Gamma}\, \equiv  (\rho_+ + \rho_-)(1 + \vert \lambda \vert^2)$, $\varphi \equiv 
\frac{t_{\uparrow}}{\vert t_{\uparrow}\vert}$, 
$\rho_{\pm} \equiv  \sqrt{\frac{\Delta }{ 2}} \, \, \frac{1 }{ \sqrt{{\cal V}}}\,
 \frac{1 }{ \sqrt{1 \mp \eta}}$, 
${\cal V} \, \equiv  eV - \Delta$ and $\eta \, \equiv \frac{E }{ {\cal V}}$.
It is straightforward to calculate $P_{\uparrow\downarrow}$ by injecting Eq. (\ref{cont_mbs}) into Eq. (\ref{spincor_def}). Thus, we get the analytical expression for the spin current correlations
\begin{equation}
P_{\uparrow\downarrow} \, = \, - 4 \Bigl(\frac{\vert \lambda \vert }{ 1 + \vert \lambda \vert^2}\Bigr)^2
 \Bigl\lbrack 3 I_+ - I_- \Bigr\rbrack \,\frac{e^2 }{ h} (eV - \Delta),
\end{equation}
with $I_+ \, = \, \int_0^1 \frac{1 - x^2 }{ (1 + \sqrt{1 - x^2})^2} \, dx \, = \, \frac{10 }{ 3} - \pi$ and
$I_- \, = \, \int_0^1 \frac{\sqrt{1 - x^2} }{ (1 + \sqrt{1 - x^2})^2} \, dx \, = \, \frac{\pi }{ 2}
 - \frac{4 }{ 3}$.
For $\lambda=1$, the average current $\langle I_1 \rangle$ is 
\begin{equation}
\langle I_1 \rangle=\frac{2e}{h}(4-\pi)(eV-\Delta).
\end{equation} 
 Combining the two previous equations leads to the final result for the spin current correlations in the case of an MBS detection
\begin{eqnarray}
F_{\uparrow\downarrow} \, = \, \frac{P_{\uparrow\downarrow}}{e \langle I_{1}\rangle} \, =
\,- \frac{68 - 21 \pi }{ 48 - 12 \pi} \simeq \, -0.197.
\end{eqnarray}

\subsection{Andreev Bound State case}

In the case of an ABS, the ${\hat r}$ matrix is
\begin{eqnarray}
{\hat r}_{ABS} \, = \, 
\begin{pmatrix}
{\tilde r}^{ee}_{\uparrow\uparrow} & 0 & 0 &{\tilde r}^{he}_{\downarrow\uparrow}    \cr
0 & {\tilde r}^{ee}_{\downarrow\downarrow} &{\tilde r}^{he}_{\uparrow\downarrow} & 0 \cr
0 &   {\tilde r}^{eh}_{\downarrow\uparrow}  & {\tilde r}^{hh}_{\uparrow\uparrow} &0 \cr

 {\tilde r}^{eh}_{\uparrow\downarrow}  & 0 & 0 & {\tilde r}^{hh}_{\downarrow\downarrow}
\end{pmatrix},
\end{eqnarray}
with
${\tilde r}^{ee}_{\uparrow\uparrow} \, = \, \frac{\rho_- \vert \lambda^2 \vert - \rho_+}{ \rho_+ + \rho_- \vert \lambda \vert^2}$,
${\tilde r}^{ee}_{\downarrow\downarrow} \, = \, 
\frac{\rho_- - \rho_+ \vert \lambda \vert^2 }{ \rho_+ \vert \lambda \vert^2 + \rho_-}$,
${\tilde r}^{he}_{\uparrow\downarrow} \, = \, {\tilde r}^{eh \, *}_{\downarrow\uparrow} \, = \, 
- 2 \frac{\sqrt{\rho_+ \rho_- } \lambda^* \varphi^{-2} }{ \rho_+ \lambda^2 + \rho_-}$,
and ${\tilde r}^{he}_{\downarrow\uparrow} \, = \, {\tilde r}^{eh \, *}_{\uparrow\downarrow} \, = \, 
- 2 \lambda^* \varphi^{-2} \frac{\sqrt{\rho_+ \rho_-}  }{ \rho_+ + \rho_- \lambda^2}$.
Because the usual Andreev reflection occurs with a spin flip and the specular reflection does not, the formula for $P_{\uparrow\downarrow}$ simplifies and gives us
\begin{equation}
P_{\uparrow\downarrow} \, = \, \frac{e^2 }{ h}\,
\int_0^{eV - \Delta}
\Bigl\lbrack | {\tilde r}^{ee}_{\uparrow\uparrow}|^2 \, | {\tilde r}^{he}_{\downarrow\uparrow}|^2 + 
| {\tilde r}^{ee}_{\downarrow\downarrow}|^2 \,
 | {\tilde r}^{he}_{\uparrow\downarrow}|^2 \Bigr\rbrack
 dE,
\end{equation}
which after replacing all the components of the reflection matrix leads to
\begin{equation}
P_{\uparrow\downarrow} \, = \, 
\frac{e^2 }{ h} (eV - \Delta) 
 \bigl\lbrack \kappa(\lambda) + \kappa(\lambda^{-1})\bigr\rbrack,
\end{equation}
with 
$\kappa(\lambda) \, = \, 
\int_0^1 4 \vert \lambda \vert ^2 s (\vert \lambda \vert^2 -s)^2 (\vert \lambda \vert^2 + s)^{-4} dx$ where 
$s=\sqrt{\frac{1-x }{ 1+x}}$. For the spin independent tunneling case $\lambda=1$, the current flowing through the ABS is twice the MBS case, namely, $\langle I_{{\rm 1}} \rangle \, = \, \frac{4e }{ h}  \,(4 - \pi)(eV - \Delta)$, leading to 
\begin{eqnarray}
F_{\uparrow\downarrow} \,  = \, \frac{9 \pi - 28 }{ 24 - 6 \pi} \simeq \, 0.053.
\end{eqnarray}


\begin{thebibliography}{99}
\bibitem{footnoteta} In this work, the tunneling amplitudes are taken to be real for the sake of simplicity. However, we have checked that our results are robust when taking complex amplitudes \cite{Nilssonetal} (see appendix B).
\end{thebibliography}
\end{document}